# A Bell Theorem Without Inequalities for Two Particles, Using Efficient Detectors

by

Daniel M. Greenberger
*City College of New York, New York, NY 10031*

Michael Horne
*Stonehill College, Easton, MA 02357*

Anton Zeilinger
*Institute for Experimental Physics, U. of Vienna, Vienna A-1090*

## Abstract

We consider an entangled two-particle state that is produced from two independent down-conversion laser sources by the process of "entanglement-swapping", so that the particles have never met. We prove a Greenberger-Horne-Zeilinger (GHZ) type theorem, showing that any deterministic, local, realistic theory is inconsistent with the quantum mechanical perfect correlations for such a state. This theorem holds for individual events, with no inequalities, for detectors of 100% efficiency.

## 1. Introduction

We would like to produce a Bell's Theorem[1] for two entangled particles that uses a Greenberger-Horne-Zeilinger (GHZ)-type argument[2]. The argument applies to the case where the two particles have a perfect correlation, meaning that if one knows the outcome of a measurement on one of them, one can predict the outcome of a corresponding measurement on the other with absolute certainty, so that an Einstein-Podolsky-Rosen (EPR) element of reality[3] exists. Another feature of the argument is that it involves no inequalities, and discusses only perfectly correlated states. Because of this, one does not have to integrate over the internal hidden variables, and one can deal directly with the hidden variable models, and show that they cannot reproduce the quantum mechanical results.

One might be skeptical that such an argument can exist, since in the usual Bohm-Bell analysis[4] for two particles in a singlet state, one can reproduce the perfect correlations with a simple classical model[5]. But that is because of the way that the singlet state is usually produced, namely through a single decay event that results in the simultaneous creation of both the particles.

However, a different way to produce two-particle entangled states has recently been successfully shown to work, namely the method of "entanglement-swapping".[6] In this method, two pairs of particles, each pair in a singlet state, are independently produced. Then one catches one particle of each pair simultaneously (which correlates them into what we call a "cross-entangled" state). This automatically correlates the other two particles, which have never met, into an entangled state. This is the "entanglement-swapped" state. This entanglement-swapped state is quantum-mechanically like any other entangled two-particle state, but from the point of view of classical local reality, it is a very problematical state, since two independently produced particles that have never met are suddenly thrown into an entanglement, even though they have no shared history. We will show that the situation is strange enough so that one cannot reproduce the quantum perfect correlations of the entanglement-swapped state with a classical, deterministic theory. While the original experiment[6] detected the cross-entangled pair of particles in a singlet state, in principle we can detect any of the four Bell states and our argument uses all four of the Bell states[7], rather than merely the singlet state.

Others have produced arguments very similar to ours, in a different context (see for example the very cleanly written papers of Hardy[8,9], Cabello[10,11], Aravind[12], Chen[13,14], and Pavicic and Summhammer[15]), rather than in a direct Bell-type experiment. In their arguments the analysis is quantum mechanical, and they do not directly analyze hidden variable models. Of the papers we quote, only Hardy does not use any inequalities. But in his discussion he uses results that "sometimes" occur, so they do not keep track of the individual hidden variables. His results also do not pertain to the perfect correlations of completely entangled states, and the experiments confirming his results[9] use the "fair-sampling" assumption. His paper also does not use entanglement swapping. So his results are in some sense complementary to ours. In the other papers, even though two down-conversions are used to make the four photons, in the quantum analysis one does not have to worry about whether they are produced by two independent lasers, or by a split beam from one laser. But in our classical analysis, it is very important that the two sets of hidden variables, one produced by each down-conversion, are inherently independent, and so they must be produced by two independent laser sources. There is a large difference, classically, between one or two totally independent sources, although it does not show up in a discussion involving 100% efficient detectors, where we also do not need to exploit all the symmetries available classically from the independence of the two-source state we use. Elsewhere[16] (referred to as paper B), we shall use the total independence of the sources to show that one can rule out realistic local theories even if one uses detectors of low efficiency.

Our papers differ also in kind from the papers we quote above in that those papers are interested in showing that an experiment can rule out various types of theories built on local realism. We are interested in showing that the local, realistic theories themselves are internally inconsistent, in the sense that they cannot be constructed to reproduce the quantum results for all perfect correlations. These realistic theories are inconsistent and cannot reproduce the quantum results, even in just the perfect correlation cases, which are where the EPR elements of reality argument holds. All individual values of the hidden variables yield definite EPR elements of reality which, taken together, produce results that cannot be fitted together to explain the quantum results for perfect correlations. These inconsistent results occur at the level of the individual variables, and do not have to be integrated over. So there is no need for "random sampling" type assumptions.

In this paper we consider only 100% efficient detectors. In both papers we are concerned with indivual values of the hidden variables, and in paper B we show that such realistic theories are inconsistent even for arbitrarily low efficiencies, provided the theories are "robust", which we will explain in detail in paper B. (Robustness requires that even for a very low efficiency realistic model, one can detect some events at every polarization angle setting. Otherwise there will be angles at which nothing can ever be observed. One cannot prove the inconsistency of a theory that predicts that no events are ever observed! We consider robustness to be a reasonable requirement.) So as long as one is willing to concede that quantum mechanics works, and that all the perfect correlation cases predicted by quantum theory hold true in the laboratory, then because these realistic theories are inconsistent, there is no experiment that needs to be done to rule them out. They are self-defeating.

Some people would argue that we (and refs. (10-15)) do not have a true two-particle state since we start with a four-particle state and reduce it by subsequent measurements. It is true that one needs all four particles to prove the existence of the various elements of reality present in the two-particle state. But once this is done one can perform and analyze EPR experiments with this two-particle state. An argument in defense of calling it a two-particle state is presented in paper B.

Recently, a paper of Broadbent and Méthot argues[17] that one can explain entanglement swapping experiments by hidden variables. Their argument does not apply to our situation, as is explained in Paper B, in footnote (14).

In this paper, we first describe the entanglement swapping experiment we will be analyzing, and examine it quantum-mechanically, noting all the perfect correlation cases that arise. Then we show how a hidden variable experiment will describe the same experiment, and derive a formula, eqs. (15) and (17), that describes all the perfect correlation cases from a local, realistic point of view. Finally, we show that this formula is inconsistent, so that no such hidden variable model can describe the perfect correlations, in the case of 100% efficient detectors. The case of inefficient detectors is more involved, and requires more elaborate machinery than we develop in this paper, which will be presented in paper B.

## 2. A Quantum Analysis of the Experiment

The Bell States of a two-particle system are a particular set of four orthogonal entangled states that form a complete set of states for the system. For a two-photon system they are

$$
\begin{aligned}
\left|\phi^+\right\rangle &= \tfrac{1}{\sqrt{2}}(\left|H_1\right\rangle\left|H_2\right\rangle + \left|V_1\right\rangle\left|V_2\right\rangle), \\
\left|\phi^-\right\rangle &= \tfrac{1}{\sqrt{2}}(\left|H_1\right\rangle\left|H_2\right\rangle - \left|V_1\right\rangle\left|V_2\right\rangle), \\
\left|\psi^+\right\rangle &= \tfrac{1}{\sqrt{2}}(\left|H_1\right\rangle\left|V_2\right\rangle + \left|V_1\right\rangle\left|H_2\right\rangle), \\
\left|\psi^-\right\rangle &= \tfrac{1}{\sqrt{2}}(\left|H_1\right\rangle\left|V_2\right\rangle - \left|V_1\right\rangle\left|H_2\right\rangle).
\end{aligned}
\tag{1}
$$

Here the subscripts 1,2 refer to two different momentum states for the different photons. The notations $\left|\phi^\pm\right\rangle, \left|\psi^\pm\right\rangle$ in eq. (1) represent the conventional labeling of each of these states. With present technology, by making suitable unitary transformations between the four Bell states, one can detect any two of the four states.

One further operation we shall need is that of rotating the polarizations of each of our photons. This is given by the equations

$$
\begin{aligned}
R(\varphi)\left|H\right\rangle &= \left|H\right\rangle\cos\varphi + \left|V\right\rangle\sin\varphi, \\
R(\varphi)\left|V\right\rangle &= \left|V\right\rangle\cos\varphi - \left|H\right\rangle\sin\varphi.
\end{aligned}
\tag{2}
$$

The situation we are going to describe is based on an experiment that was recently performed by the Zeilinger group[6], in which they swapped the entanglement of two photons as mentioned in the introduction, and as depicted in Fig. (1). In this experiment, two independent pairs of photons are created, each in the photon equivalent of a singlet state, $\tfrac{1}{\sqrt{2}}(H_1V_2 - V_1H_2) = \left|\psi^-\right\rangle$, and the polarizations of the four photons are independently rotated, through the angles $\varphi_1$, $\varphi_2$, $\varphi_3$, and $\varphi_4$, as in eq. (2)).

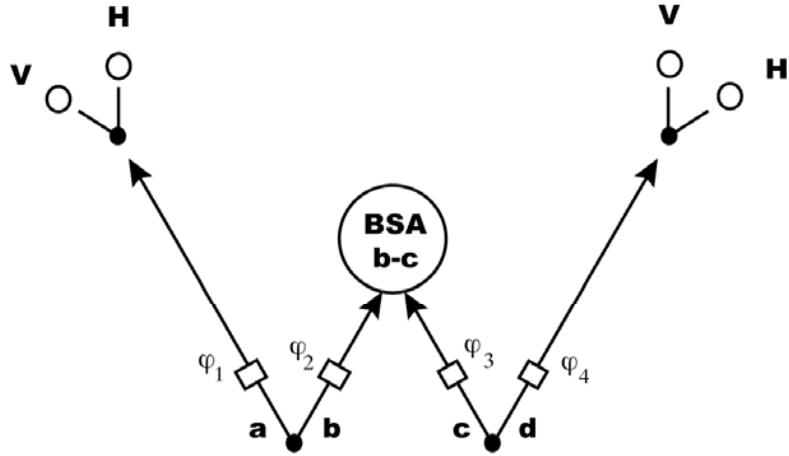

## Fig. (1)

Figure (1).  Schematic Diagram of the Creation of the Two-Particle State
In this experiment there are two down-conversions, one creating the pair of photons $a$-$b$, and the other the pair $c$-$d$.  Each photon undergoes a rotation through the angle $\varphi_i$, and particles $b$ and $c$ enter a Bell-state-analyzer (BSA), which will annihilate them while detecting which Bell state they were in.  If the angles $\varphi_i$ are set properly, as one of the perfect correlation cases, this process forces the particles a and d into a two-particle Bell state.  In the actual experiment the Bell state of $a$ and $d$ is not determined, only their polarizations, but this is sufficient to rule out locally realistic, deterministic theories as an explanation of their observed properties.

The initial state of the system produced by the two independent lasers and two independent down-conversions[18] is a product of two singlet states, photons $a$ and $b$ produced by one laser, and $c$ and $d$ produced by the other,

$$|\psi_0\rangle = \tfrac{1}{2}(|H\rangle_a|V\rangle_b - |V\rangle_a|H\rangle_b)(|H\rangle_c|V\rangle_d - |V\rangle_c|H\rangle_d),$$ (3)

We call this the "Volkswagen state" (VW state), from the shape of Fig. (1).  Next, the polarization of each of the photons $a$, $b$, $c$, and $d$, gets rotated through their respective angles, $\varphi_i$.  The subsequent experiment combines photons $b$ and $c$ at a Bell state analyzer (BSA)[19] and also detects photons $a$ and $d$ separately, as well as their polarizations.  It is thus appropriate to rewrite eq. (3) in terms of the Bell states of particles $b$ and $c$, and the Bell states of particles $a$ and $d$, after the polarization of each of the photons has been rotated.  The resulting wave function, $|\psi_1\rangle$, after some algebra, is

$$\begin{aligned}
|\psi_I\rangle = \tfrac{1}{2}\Big\{ & \left|\phi_{bc}^+\right\rangle\left[-\left|\phi_{ad}^+\right\rangle\cos\xi + \left|\psi_{ad}^-\right\rangle\sin\xi\right] \\
& + \left|\psi_{bc}^-\right\rangle\left[-\left|\psi_{ad}^-\right\rangle\cos\xi - \left|\phi_{ad}^+\right\rangle\sin\xi\right] \\
& + \left|\phi_{bc}^-\right\rangle\left[+\left|\phi_{ad}^-\right\rangle\cos\eta + \left|\psi_{ad}^+\right\rangle\sin\eta\right] \\
& + \left|\psi_{bc}^+\right\rangle\left[+\left|\psi_{ad}^+\right\rangle\cos\eta - \left|\phi_{ad}^-\right\rangle\sin\eta\right]\Big\},
\end{aligned}$$

$$\xi = ((\varphi_1 - \varphi_2) + (\varphi_3 - \varphi_4)),$$
$$\eta = ((\varphi_1 - \varphi_2) - (\varphi_3 - \varphi_4)).$$

$(4)$

As they approach the BSA, the two particles $b$ and $c$ are in a superposition of all four Bell states, according to eq. (4). However, in the original Zeilinger experiment, only a beam-splitter was used, instead of a BSA, and so only the singlet state $\left|\psi^-\right\rangle$ was detected[6], and if the angle $\xi = 0$, then by eq. (4) particles $a$ and $d$ were also thrown into a singlet state.

One can see from Fig. (1) that the original experiment can also be conceived of as a sort of teleportation[20]. Consider one of the incident entangled pairs of photons, say $a$ and $b$. Then by catching photons $b$ and $c$ simultaneously and making a Bell-state measurement on them, one has "teleported" the entanglement of photons $a$ and $b$ into that of photons $a$ and $d$. Normally, one must telegraph classical information from the detectors at the BSA to a device in the path of particle $d$, in order to allow it to be appropriately manipulated, so that it will be in the proper configuration. However, in the case of singlets, the appropriate manipulation is the unit operator. The only classical information that has to be transmitted is the fact that the Bell state measurement has been successfully made.

The special feature of such a teleportation is that the resulting entanglement is between two photons that were created independently and that have never met. Because of the symmetry of the situation, one can instead consider the entanglement of photons $d$ and $c$ to have been teleported to that of photons $d$ and $a$. Another interesting aspect of this situation is that because it is entangled, particle $b$ cannot be said to have a specific state that has been teleported. Rather, it is the entanglement itself that has been "teleported". In this experiment, only the singlet Bell state was detected, yielding only the term $\left|\psi_{bc}^-\right\rangle\left|\psi_{ad}^-\right\rangle$ in eq. (4), and the particles $a$ and $d$ were entangled with high fidelity, when $\xi = 0$. In the experiment, the phase shifter $\varphi_3$ in beam $c$ was only used as a fiduciary phase, in order to determine the zero point for the other three phases.

With a fully functional BSA, all of the four Bell states would be detectable. Then eq. (4) shows that, once the BSA result is registered, particles $a$ and $d$ are in general thrown into a superposition of the Bell states, and with a suitable choice of the $\varphi_i$, into a very specific Bell state. In what follows, we will assume a full BSA.

### 3. Analysis of the Arrangement in Terms of Elements of Reality

We shall start by establishing the elements of reality implied by the perfect correlations in experiments using the VW state, restricting ourselves in this paper to 100% efficient detectors, as originally envisioned by EPR in their analysis. First, we consider the elements of reality directly connected to the Bell states of particles $b$ and $c$, and to the Bell states of the particles $a$ and $d$ as well.

Consider the arrangement, shown in Fig. (2), which is almost that of Fig. (1), but which is more symmetrical and more general.  In Fig. (2) we have included a BSA that can detect the Bell state of particles *a* and *d*, as well as the one that works with particles *b* and *c*.  We have also drawn one of our down-conversion pairs coming from the bottom and one from the top.  If one makes both these Bell state measurements at the same time, then according to eq. (4), the results will be strongly correlated.  In fact, according to eq. (4), if one finds that the Bell state for particles *b* and *c* is either $\phi^-$ or $\psi^+$, then one will invariably find that the Bell state for particles *a* and *d* will also be $\phi^-$ or $\psi^+$.  This statement will be true regardless of the values of any of the angles $\varphi_i$.  Similarly, the Bell states $\phi^+$ and $\psi^-$ for particles *b* and *c* are invariably coupled with the Bell states $\phi^+$ and $\psi^-$ for particles *a* and *d*, again regardless of the values of the $\varphi_i$.

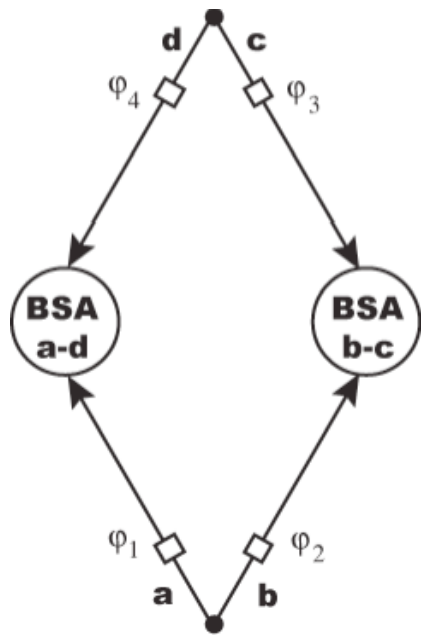

**Fig. (2)**

Figure (2).  <u>Bell-State Determination of Both Sets of Particles</u>
This experiment is similar to that in Fig. (1), except that the Bell states of both sets of particles, *b-c*, and *a-d*, are determined at the Bell-state-analyzers (BSA).  One set of particles is shown entering from the top, to explicitly show the symmetry of the situation.  When the particles are created, in a deterministic treatment, either the situation of Fig. (1) or that of Fig. (2) can be set up by the experimenter before the particles reach their respective BSA's, and so both possibilities must be forseen in the original instructions to the particles.

These results show the existence of an EPR element of reality associated with these Bell state measurements.  Recall the EPR definition of an element of reality[3]:  *If, without in any way disturbing a system, we can predict with certainty (i.e., with probability equal to unity) the value of a physical quantity, then there exists an element of physical reality corresponding to this physical quantity*.  In our case, if one measures the Bell state of particles *b* and *c* to be one of the outcomes $\phi_{bc}^-, \psi_{bc}^+$, then one can predict with 100% certainty that if one also measures the Bell state of particles $a$ and $d$, it too will turn out to be $\phi_{bc}^-, \psi_{bc}^+$, which one can determine even if the first measurement is performed far away from the second one, without the results in any way disturbing particles a and d.  A similar situation occurs if the result of the first measurement turns out to be $\phi_{bc}^+, \psi_{bc}^-$.

EPR also set up a requirement for a theory to be complete[3]: *Every element of physical reality must have a counterpart in the physical theory*.  A consequence of completeness here is the existence of a function[21]

$$\kappa = \kappa(\varphi_2, \varphi_3, \lambda_1, \lambda_4), \tag{5}$$

for the outcome of the Bell measurement for particles $b$ and $c$, whose values equal +1 for the Bell states $\phi^+$ and $\psi^-$, and -1 for the Bell states $\phi^-$ and $\psi^+$.  Here, $\kappa$ can only depend on the angles $\varphi_2$ and $\varphi_3$, as a look at Fig. (2) shows that the two particles meeting at the BSA have only encountered these two angles in their flight to the analyzer.  The function $\kappa$ can also depend on any hidden variables, collectively noted by $\lambda_1$, that were defined for the particles $a$ and $b$, when they were created together, and on those, $\lambda_4$, for the particles $c$ and $d$, when they were created together.

These hidden variables arise from the fact that in a deterministic local, realistic model, all characteristics of the various particles, such as their polarizations, and any correlations, must be defined for the particles when they are created.  They may evolve over time into new values, and change values as they meet various other particles, or detectors along the way, but these are deterministic consequences unfolding from the initial assignments and the equations of motion, in such a way as to produce a continuous local explanation of all causes and effects in the history of the particle as it moves along.  We assign these values when the particles are created, and denote them by the $\lambda$'s.

Similarly, there exists a function

$$\kappa' = \kappa'(\varphi_1, \varphi_4, \lambda_1, \lambda_4), \tag{6}$$

for the outcome of the measurements of the Bell states of the particles $a$ and $d$.  Again, the function $\kappa'$ takes on the value +1 for Bell states $\phi^+$ and $\psi^-$, and -1 for Bell states $\phi^-$ and $\psi^+$.

Then the statement of the perfect correlation between the two BSA measurements in Fig. (2) is given by the equation

$$\kappa(\varphi_2, \varphi_3, \lambda_1, \lambda_4)\, \kappa'(\varphi_1, \varphi_4, \lambda_1, \lambda_4) = 1,$$
$$\kappa(\varphi_2, \varphi_3, \lambda_1, \lambda_4) = \kappa'(\varphi_1, \varphi_4, \lambda_1, \lambda_4). \tag{7}$$

Since the left hand side of this equation only depends on the angles $\varphi_2$ and $\varphi_3$, while the right hand side only depends on the angles $\varphi_1$ and $\varphi_4$, it follows that neither side depends upon any of the angles $\varphi_i$.  This could also have been seen from the original statement that the perfect correlations that lead to eq. (7) exist independently of any of these angles, which follow from the form of eq. (4).  So we can restate eq. (7) as

$$\kappa(\lambda_1, \lambda_4) = \kappa'(\lambda_1, \lambda_4),$$
$$\text{independently of the } \varphi_i. \tag{8}$$

Eq. (4) also shows that for certain sets of the $\varphi_i$, there are additional perfect correlations, and additional elements of reality.  If the angle $\xi = \varphi_1 - \varphi_2 + (\varphi_3 - \varphi_4) = 0, \pm\pi$, in eq. (4), then when the Bell state of particles $b$ and $c$ turns out to be $\phi_{bc}^+$ $[\psi_{bc}^-]$, the Bell state of particles $a$ and $d$ will also be $\phi_{ad}^+$ $[\psi_{ad}^-]$.  So in this case, there is an exact correlation of the individual Bell states, not merely a partial correlation to a set of two of them.  Similarly, if $\xi = \pm\pi/2$, then the opposite exact correlation will hold, namely when the Bell state of particles $b$ and $c$ turns out to be $\phi_{bc}^+$ [$\psi_{bc}^-$], the Bell state of particles $a$ and $d$ will be $\psi_{ad}^-$[$\phi_{ad}^+$]. These are the perfect correlations when $\kappa = +1$.

When $\kappa = -1$, there is a similar set of perfect correlations. In this case, according to eq. (4), if the angle $\eta = \varphi_1 - \varphi_2 - (\varphi_3 - \varphi_4) = 0, \pm\pi$, then, when the Bell state of particles $b$ and $c$ turns out to be $\phi_{bc}^-$ [$\psi_{bc}^+$], the Bell state of particles $a$ and $d$ will also be $\phi_{ad}^-$ [$\psi_{ad}^+$]. Similarly, if $\eta = \pm\pi/2$, then the opposite exact correlation will hold, namely when the Bell state of particles $b$ and $c$ turns out to be $\phi_{bc}^-$ [$\psi_{bc}^+$], the Bell state of particles $a$ and $d$ will be $\psi_{ad}^+$ [$\phi_{ad}^-$]. These cases represent the perfect correlations when $\kappa = -1$.

Additional information can be gleaned from these perfect correlations by considering the polarization states of the particles. We give the numerical value +1 to an H polarization, and -1 to a V polarization. Then a Bell state can be labelled by the product of the polarization of its two particles. That is, if the Bell state contains the polarizations HH and VV, the product is +1, (see eq. (1)), while if it contains the polarizations HV and VH, the product is -1. Because of the perfect correlations summarized in the above paragraph by the existence of $\kappa$ and $\kappa'$, there exists a function

$$F = F_{\kappa(\lambda_1, \lambda_4)}(\varphi_2, \varphi_3, \lambda_1, \lambda_4) = \pm 1, \tag{9}$$

that represents the polarization product of the Bell state of particles $b$ and $c$. Here F $= +1$ for HH or VV, representing the states $\phi_{bc}^+$ and $\phi_{bc}^-$, and F $= -1$ for HV and VH, representing the states $\psi_{bc}^+$ and $\psi_{bc}^-$. F also depends on the value of $\kappa$, and so, since it includes both the polarization product and the $\kappa$ value of the state, it uniquely represents the Bell state of the particles $b$ and $c$. Explicitly,

$$
\begin{aligned}
F_1 &= +1, & \textit{refers to} & & \phi_{bc}^+, \\
F_1 &= -1, & " & & \psi_{bc}^-, \\
F_{-1} &= +1, & " & & \psi_{bc}^+, \\
F_{-1} &= -1, & " & & \phi_{bc}^-.
\end{aligned}
\tag{10}
$$

One could alternatively (and equivalently) characterize F as a function that can take on four values, rather than two, given namely by the values of $\kappa$ and the polarization. However the way we have done it will be convenient for characterizing the perfect correlations in our experiments.

In the same way as we have done for the two particles $b$ and $c$, we can uniquely characterize the Bell state of the particles $a$ and $d$, by introducing the function

$$G = G_{\kappa'(\lambda_1, \lambda_4)}(\varphi_1, \varphi_4, \lambda_1, \lambda_4) = \pm 1. \tag{11}$$

Once again, the values G $= \pm 1$ give the polarization product of the Bell state, and $\kappa'$ characterizes which subset, as before.

These results show that the full ensemble of states described by the parameters $(\lambda_1, \lambda_4)$ breaks into two distinct, non-overlapping sub-ensembles. Half of the sets $(\lambda_1, \lambda_4)$ work with one value of $\kappa$, and half work with the other value. So eq. (8) implies that once one has a value $\pm 1$ for $\kappa$, it will not change, regardless of the $\varphi_i$. However, for each pair of the $(\lambda_1, \lambda_4)$, all values of the $\varphi_i$ do occur, since they are determined by the experimenter (possibly as late as while the beam is traveling toward the Bell state analyzer, or detectors), whereas the $\lambda$'s are chosen by nature when the photons are created.

All of the perfect correlations that we have discussed in the arrangement of Fig. (2) are summed up by the conditions

$$F_{\kappa(\lambda_1,\lambda_4)}(\varphi_2,\varphi_3,\lambda_1,\lambda_4)G_{\kappa(\lambda_1,\lambda_4)}(\varphi_1,\varphi_4,\lambda_1,\lambda_4)=1, \quad \zeta_\kappa=0,\pm\pi,$$

$$F_{\kappa(\lambda_1,\lambda_4)}(\varphi_2,\varphi_3,\lambda_1,\lambda_4)G_{\kappa(\lambda_1,\lambda_4)}(\varphi_1,\varphi_4,\lambda_1,\lambda_4)=-1, \quad \zeta_\kappa=\pm\tfrac{\pi}{2}, \qquad (12)$$

$$\zeta_\kappa=\varphi_1-\varphi_2+\kappa(\lambda_1,\lambda_4)(\varphi_3-\varphi_4).$$

Here, we have dropped the prime on κ', since κ' = κ (see eq. (8)). Also we have introduced the angle $\zeta_\kappa$, which replaces both ξ and η, $(\zeta_+=\xi,\ \zeta_-=\eta)$.

This analysis can be extended to our actual experiment, employing the VW state of Fig. (1), where we will not combine the particles *a* and *d* into a Bell State, but will measure their polarizations separately. According to eq. (4), it is still true that the function F exists in this case, as the particles *b* and *c* cannot tell if we are going to measure particles *a* and *d* separately, or combine them into a Bell State. Particles *a* and *d* each have an element of reality associated with their polarization since, if we measure the Bell state of particles *b* and *c* and then measure the polarization of particle *d*, we can uniquely predict the polarization of particle *a* without interacting with it in any way, provided the angle $\zeta_\kappa=0,\pm\pi/2,\pm\pi$. A similar argument holds for particle *d*. Thus there exist functions A and D such that

$$A=A(\varphi_1,\lambda_1)=\pm 1,$$

$$D=D(\varphi_4,\lambda_4)=\pm 1, \qquad (13)$$

where, as before, A = +1 for polarization H, and A= -1 for polarization V. In this case, there is no operational significance to the function κ'. Any value of $\lambda_1$ is compatible with both values of κ', depending on the value of $\lambda_4$. (This argument shows the existence of the function A($\varphi_1,\lambda_1$) within the context of our experiment. But this also follows directly from the fact that particles *a* and *b* are created in a singlet state, and so by setting $\varphi_2=\varphi_1$, and detecting particle *b* before it reaches the BSA, one can uniquely predict the polarization of *a* from the outcome for *b*.)

Now, we have all the elements of reality needed for a local-reality account of the arrangement of Fig. (1), where the particles *a* and *d* do not meet at a BSA, but go directly and separately to polarization analyzers. Agreement with perfect correlations in the quantum state of eq. (4) imposes the conditions

$$A(\varphi_1,\lambda_1)F_{\kappa(\lambda_1,\lambda_4)}(\varphi_2,\varphi_3,\lambda_1,\lambda_4)D(\varphi_4,\lambda_4)=1, \quad \zeta_\kappa=0,\pm\pi,$$

$$A(\varphi_1,\lambda_1)F_{\kappa(\lambda_1,\lambda_4)}(\varphi_2,\varphi_3,\lambda_1,\lambda_4)D(\varphi_4,\lambda_4)=-1, \quad \zeta_\kappa=\pm\tfrac{\pi}{2}, \qquad (14)$$

$$\zeta_\kappa=\varphi_1-\varphi_2+\kappa(\lambda_1,\lambda_4)(\varphi_3-\varphi_4),$$

on the product of the polarizations, given by the elements of reality functions A, $F_\kappa$, D, and κ.

From a quantum-mechanical point of view the particles *a* and *d* are entangled, once the particles *b* and *c* are detected by the BSA. From a classical, realistic point of view all the correlations between the particles must be contained in eq. (14). Photon *a* is coupled to photon *b* at its creation, and any instructions to photon *a* were built in when the state was created. The situation is the same with photons *c* and *d*, which were created independently of photons *a* and *b*. The instructions to photons *a* and *d* have no input from whatever happens when photons *b* and *c* meet at the Bell state analyzer, BSA, (and indeed, in the actual Zeilinger experiment[6] we mentioned earlier, photons *a* and *d* were counted before any conceivable information could propagate to them at the speed of light from the beam-splitter through which particles *b* and *c* passed).

In fact, since both pairs were independently created, the instructions to photons *a* and *d* are completely uncorrelated. Nevertheless, photons *a* and *d* become entangled when both photons *b* and *c* are counted at the BSA. Thus, from a classical point of view, extra information is generated at the BSA when particles *b* and *c* meet, and it is not obvious *a priori* how strongly the original instructions can incorporate this information to correlate the particles *a* and *d*. However, so long as we restrict ourselves to 100% efficient counters, we shall see that this effect is minimal.

### 4. The Inconsistency of Local, Realistic Models

All of our counters are assumed to be 100% efficient, so that for any value of any of the $\lambda$'s, and any of the $\varphi$'s, all of our functions, $\kappa$, A, F, and D, are defined and take the value $\pm 1$. In the perfect correlation conditions, eq. (14), we take the special case $\varphi_1 = \varphi_2$, $\varphi_3 = \varphi_4$, so that $\zeta_\kappa = 0$ for either value of $\kappa$, leading to

$$A(\varphi_2, \lambda_1) F_{\kappa(\lambda_1, \lambda_4)}(\varphi_2, \varphi_3, \lambda_1, \lambda_4) D(\varphi_3, \lambda_4) = 1,$$
$$\zeta_\kappa = \varphi_1 - \varphi_2 + \kappa(\varphi_3 - \varphi_4) \tag{15}$$
$$= \varphi_2 - \varphi_2 + \kappa(\varphi_3 - \varphi_3) = 0.$$

Then for either value of $\kappa$,

$$F_{\kappa(\lambda_1, \lambda_4)}(\varphi_2, \varphi_3, \lambda_1, \lambda_4) = A(\varphi_2, \lambda_1) D(\varphi_3, \lambda_4). \tag{16}$$

and this reduces eq. (14) to

$$A(\varphi_1, \lambda_1) A(\varphi_2, \lambda_1) D(\varphi_3, \lambda_4) D(\varphi_4, \lambda_4) = \begin{cases} +1, & \zeta_\kappa = 0, \pm\pi, \\ -1, & \zeta_\kappa = \pm\frac{\pi}{2}. \end{cases} \tag{17}$$

One can further restrict these equations in many ways, but for the purpose of this paper it will not be necessary.

We shall show that the set of equations (17) is inconsistent. First look at the case where the variables $(\lambda_1, \lambda_4)$ are such that $k(\lambda_1, \lambda_4) = +1$. From eqs. (16), (17), if we take $\varphi_1 = \alpha$, $\varphi_2 = \alpha + \frac{\pi}{4}$, $\varphi_3 = \beta + \frac{\pi}{4}$, $\varphi_4 = \beta$, with $\alpha$ and $\beta$ arbitrary, we have

$$A(\alpha, \lambda_1) A(\alpha + \tfrac{\pi}{4}, \lambda_1) D(\beta + \tfrac{\pi}{4}, \lambda_4) D(\beta, \lambda_4) = 1,$$
$$\zeta_+ = \varphi_1 - \varphi_2 + \varphi_3 - \varphi_4 = \alpha - (\alpha + \tfrac{\pi}{4}) + (\beta + \tfrac{\pi}{4}) - \beta = 0. \tag{18}$$

If we now take the case $\varphi_1 = \alpha$, $\varphi_2 = \alpha + \frac{\pi}{4}$, $\varphi_3 = \beta$, $\varphi_4 = \beta + \frac{\pi}{4}$, we have

$$A(\alpha, \lambda_1) A(\alpha + \tfrac{\pi}{4}, \lambda_1) D(\beta, \lambda_4) D(\beta + \tfrac{\pi}{4}, \lambda_4) = -1,$$
$$\zeta_+ = \varphi_1 - \varphi_2 + \varphi_3 - \varphi_4 = \alpha - (\alpha + \tfrac{\pi}{4}) + \beta - (\beta + \tfrac{\pi}{4}) = -\tfrac{\pi}{2}. \tag{19}$$

Now multiply eq. (18) by eq. (19) to get

$$[A(\alpha, \lambda_1) A(\alpha + \tfrac{\pi}{4}, \lambda_1) D(\beta, \lambda_4) D(\beta + \tfrac{\pi}{4}, \lambda_4)]^2 = -1,$$
$$+1 = -1. \tag{20}$$

So for the case $\kappa = +1$, eqs. (17) are inconsistent.

The argument is almost identical in the case $\kappa = -1$. In that case take $(\lambda_1, \lambda_4)$ to be such that $k(\lambda_1, \lambda_4) = -1$. Then from eqs. (16), (17), take $\varphi_1 = \alpha$, $\varphi_2 = \alpha + \frac{\pi}{4}$, $\varphi_3 = \beta$, $\varphi_4 = \beta + \frac{\pi}{4}$, with $\alpha$ and $\beta$ arbitrary. Then

$$A(\alpha, \lambda_1) A(\alpha + \tfrac{\pi}{4}, \lambda_1) D(\beta, \lambda_4) D(\beta + \tfrac{\pi}{4}, \lambda_4) = +1,$$
$$\zeta_- = \varphi_1 - \varphi_2 - (\varphi_3 - \varphi_4) = \alpha - (\alpha + \tfrac{\pi}{4}) - \beta + (\beta + \tfrac{\pi}{4}) = 0. \tag{21}$$

Finally, take $\varphi_1 = \alpha$, $\varphi_2 = \alpha + \frac{\pi}{4}$, $\varphi_3 = \beta + \frac{\pi}{4}$, $\varphi_4 = \beta$, and we get

$$A(\alpha, \lambda_1)A(\alpha + \tfrac{\pi}{4}, \lambda_1)D(\beta + \tfrac{\pi}{4}, \lambda_4)D(\beta, \lambda_4) = -1,$$

$$\zeta_- = \varphi_1 - \varphi_2 - (\varphi_3 - \varphi_4) = \alpha - (\alpha + \tfrac{\pi}{4}) - (\beta + \tfrac{\pi}{4}) + \beta = -\tfrac{\pi}{2}. \qquad (22)$$

Multiplying eqs. (21) and (22) together again gives eq. (20), and so eqs. (17) are inconsistent in this case also.

Eqs. (18) to (22) follow from the definition of $\zeta_\kappa = \varphi_1 - \varphi_2 + \kappa(\varphi_3 - \varphi_4)$, so that the order of factors is very important, implying as it does, a different experiment (setting $\varphi_3$ vs. setting $\varphi_4$). Eq. (20) holds for any values of $\alpha$ and $\beta$, as we have said, and for $\kappa = \pm 1$.

We have thus proven that the set of equations, eqs. (17), and therefore eqs. (14), representing the perfect correlations in a deterministic, local, realistic, model for the entangled two-particle state described by this experiment is inconsistent when the detectors are 100% efficient, and this proof does not depend on the two down-conversion sources being totally independent, or on the elaborate factorization that will be necessary in the case of inefficient detectors, as given in paper B.

We would like to thank Prof. David Mermin for suggesting several modifications of the paper to us.

**Figure Captions**

Figure (1).  <u>Schematic Diagram of the Creation of the Two-Particle State</u>
        In this experiment there are two down-conversions, one creating the pair of photons $a$-$b$, and the other the pair $c$-$d$.  Each photon undergoes a rotation through the angle $\varphi_i$, and particles $b$ and $c$ enter a Bell-state-analyzer (BSA), which will annihilate them while detecting which Bell state they were in.  If the angles $\varphi_i$ are set properly, as one of the perfect correlation cases, this process forces the particles a and d into a two-particle Bell state.  In the  actual experiment the Bell state of $a$ and $d$ is not determined, only their polarizations, but this is sufficient to rule out locally realistic, deterministic theories as an explanation of their observed properties.

Figure (2).  <u>Bell-State Determination of Both Sets of Particles</u>
        This experiment is similar to that in Fig. (1), except that the Bell states of both sets of particles, $b$-$c$, and $a$-$d$, are determined at the Bell-state-analyzers (BSA).  One set of particles is shown entering from the top, to explicitly show the symmetry of the situation.  When the particles are created, in a deterministic treatment, either the situation of Fig. (1) or that of Fig. (2) can be set up by the experimenter before the particles reach their respective BSA's, and so both possibilities must be forseen in the original instructions to the particles.

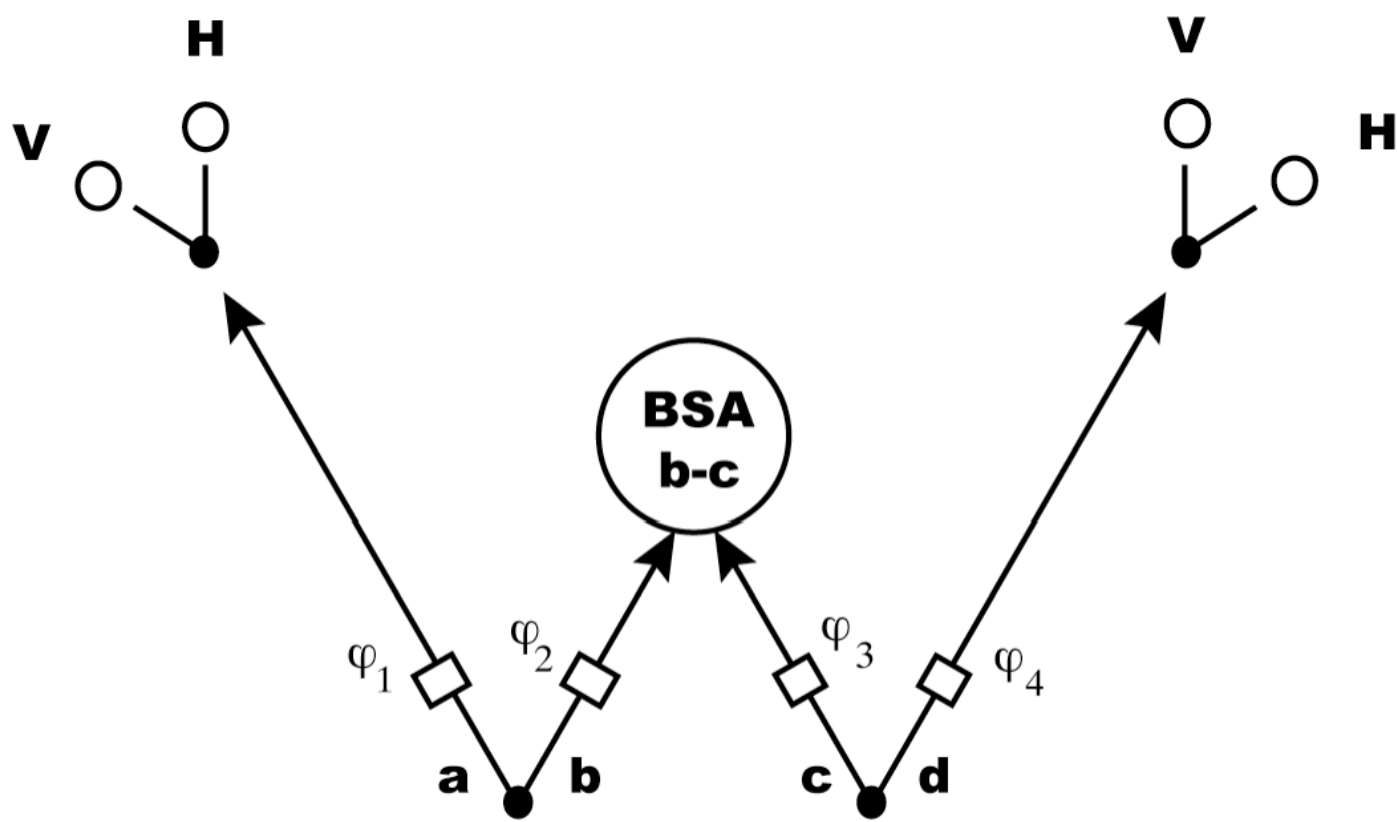

**Fig. (1)**

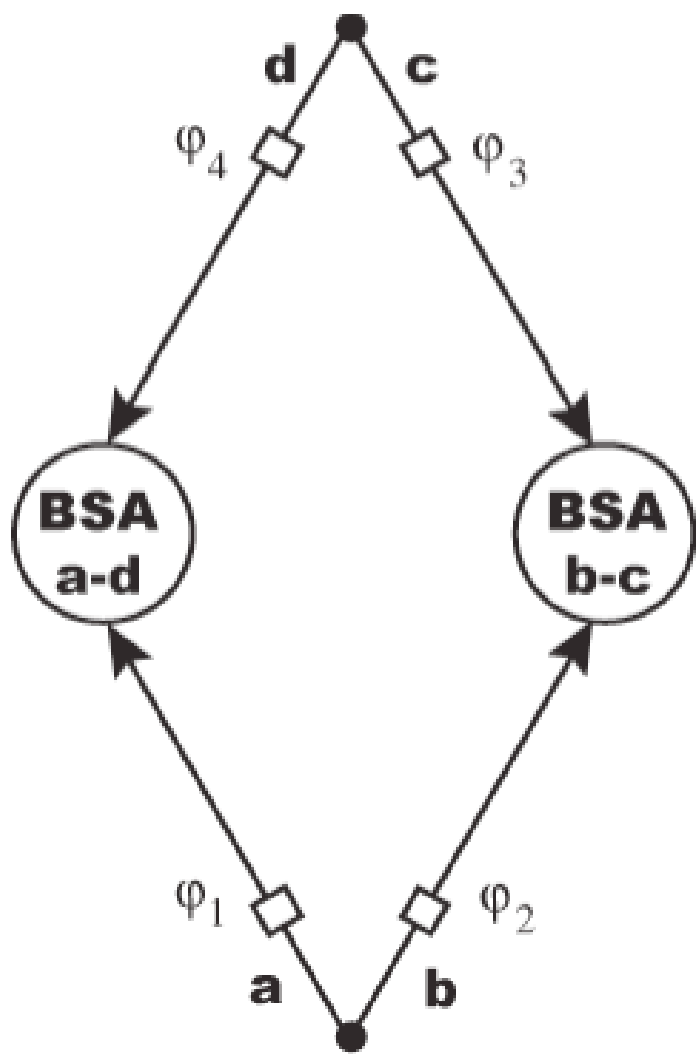

**Fig. (2)**